\documentclass[aps,prd,onecolumn,showpacs,showkeys,amsmath,amssymb]{revtex4}
\usepackage{graphicx}
\usepackage{dcolumn}
\usepackage{amsmath}
\usepackage{bm}
\begin{document}

\def\Tr{{\rm Tr}}
\def\sla#1{{\not\mathrel #1}}
\allowdisplaybreaks

\title{Pseudoscalar Meson and Decuplet Baryon Scattering Lengths}
\author{Zhan-Wei Liu}
\email{liuzhanwei@pku.edu.cn} \affiliation{Department of Physics
and State Key Laboratory of Nuclear Physics and Technology\\
Peking University, Beijing 100871, China}

\author{Yan-Rui Liu}
\email{yrliu@th.phys.titech.ac.jp} \affiliation{Department of
Physics, H-27, Tokyo Institute of Technology, Meguro, Tokyo
152-8551, Japan}

\author{Shi-Lin Zhu}
\email{zhusl@pku.edu.cn} \affiliation{Department of Physics
and State Key Laboratory of Nuclear Physics and Technology\\
and Center of High Energy Physics, Peking University, Beijing
100871, China }

\begin{abstract}
We have systematically calculated the S-wave pseudoscalar meson
and decuplet baryon scattering lengths to the third order in the
small scale expansion scheme of the heavy baryon chiral
perturbation theory. We hope the future study may test the
framework and the present predictions.
\end{abstract}



\maketitle

\pagenumbering{arabic}

\section{Introduction}\label{sec1}

The excited states of the nucleon and their strong interactions
with various hadrons play an important role in our understanding
of non-perturbative QCD. Generally speaking, their coupling with
the pseudoscalar pion, kaon and eta is governed by chiral
dynamics. The first excited state of the nucleon, $J^P=3/2^+$
$\Delta(1232)$, belongs to the lightest decuplet. It is degenerate
with the nucleon in the large $N_c$ limit
\cite{Hooft1974,Witten1979}. From the strong decay widths, one can
extract the coupling constants of the pseudoscalar meson decuplet
baryon interaction. Although it may be quite difficult to measure
their scattering lengths experimentally, unlike in the
pion-nucleon and kaon-nucleon cases
\cite{Martin1975,Koch1980,Arndt1995}, it is still interesting to
study the scattering processes theoretically.

As two well-known non-perturbative approaches of QCD at low
energy, chiral perturbation theory and lattice QCD have been
widely used in studying meson-baryon interactions
\cite{Lee1994,Mojzis1998,Kaiser2001,Scherer2003,Meng2004,Cohen2006,Bernard2008,Beane2010,Torok2010}.
With the explicitly consistent power counting scheme, the heavy
baryon chiral perturbation theory (HB$\chi$PT) is a practical and
useful tool in the study of the meson baryon scattering. The
expansion is organized with $p/\Lambda_\chi$, where $p$ represents
the momentum (mass) of the meson or the small residue momentum of
the baryon in the non-relativistic limit, and $\Lambda_\chi=4\pi
f_\pi$ is the scale of chiral symmetry breaking. Numerically it is
around the baryon mass in the chiral limit. Because of the large
mass of the strange quark, the chiral corrections from the kaon
and eta meson loops are significantly larger than that from the
pion loop, which leads to the worse convergence of the chiral
expansion in the SU(3) case.

In HB$\chi$PT, the effects of excited baryons are encoded in the
unknown low energy constants (LECs) in the Lagrangians. In the
large $N_c$ limit, one expects that the inclusion of the decuplet
baryon contribution may partially cancel the octet contribution
and be helpful to the chiral expansion. In order to investigate
the contributions of the $\Delta$, HB$\chi$PT with explicit
spin-3/2 baryons and the small-scale expansion scheme have been
developed in Refs. \cite{Hemmert1997,Hemmert1998}. Now the
expansion is organized with $\epsilon$, where $\epsilon$ means $p$
in HB$\chi$PT or the mass difference $\delta$ between $\Delta$ and
the nucleon in the chiral limit, which is widely used in the
exploration of the decuplet related processes \cite
{Hemmert1998a,Fettes2001,Zhu2001b,Zhu2002,Liu2007a}. There exists
another power counting scheme ``$\delta$ expansion'' in the
two-flavor case, where $m_\pi$ scales as $\delta^2$
\cite{Pascalutsa2003,Pascalutsa2005,Pascalutsa2005a,Pascalutsa2008}.

In our previous work \cite{Liu2007a}, we studied the decuplet
contributions to the pseudoscalar meson octet baryon scattering
lengths, which is an important observable reflecting their strong
interaction near the threshold. The inclusion of the decuplet
baryons does not improve the convergence significantly. In this
paper, we will study the scattering lengths of the pseudoscalar
mesons and the light decuplet baryons in the framework of the
SU(3) heavy baryon chiral perturbation theory with explicit
decuplet baryons. We will explore whether the small scale
expansion works well in this case.

This paper is organized as follows. We present the basic notations
and definitions in Sec. \ref{sec2}. In Sec. \ref{sec3}, we list
the chiral corrections to the threshold $T$-matrices of the meson
decuplet baryon scattering, which contains our main results. The
LECs are estimated in Sec. \ref{sec4}. Finally, we give the
numerical results and our discussions in Sec. \ref{sec5}.

\section{Lagrangians}\label{sec2}

In the heavy baryon formalism, the chiral Lagrangians at the
leading order are,
\begin{eqnarray}
  {\mathcal L}_{\phi \phi}^{(2)}&=&f^2 \Tr\left( u_\mu u^\mu+\frac{\chi_+}{4} \right), \\
  {\mathcal L}_{\phi B}^{(1)}&=& \Tr\left( \bar B (iv\cdot \partial B+[v\cdot \Gamma, B])\right)
                              +2D \Tr\left( B \{S\cdot u,B\}\right)
                              +2F \Tr\left(\bar B [S\cdot u, B] \right), \\
  {\mathcal L}_{\phi BT}^{(1)}&=& -\bar T^\mu (i v\cdot D-\delta)T_\mu
                                +{\mathcal C} (\bar T^\mu u_\mu B +\bar B u_\mu T^\mu)
                                +2{\mathcal H} \bar T^\mu S\cdot u T_\mu.\label{L1}
\end{eqnarray}
Here $f$ is the decay constant of the pseudoscalar meson in the
chiral limit, $v$ is the velocity of the baryon, $S_\mu$ is the
spin operator, and $\delta$ is the mass difference between
$\Delta$ and the nucleon. For the coupling constants,
$D+F=g_A=1.26$ where $g_A$ is the axial vector coupling constant,
$|{\cal C}|=1.5$, and $|{\cal H}|=1.9$. The $D$ and $F$ terms do
not contribute to the threshold amplitudes we calculate. The
superscript in the Lagrangians represents the corresponding chiral
order. The notations for the fields read
\begin{equation}
\Gamma_\mu = {i\over 2} [\xi^\dagger, \partial_\mu\xi],\qquad
u_\mu={i\over 2} \{\xi^\dagger, \partial_\mu \xi\},\qquad \xi =
\exp(i \phi/2f),
\end{equation}
\begin{equation}
\chi_\pm = \xi^\dagger\chi\xi^\dagger\pm\xi\chi\xi,\qquad
\chi=\mathrm{diag}(m_\pi^2,\, m_\pi^2,\, 2m_K^2-m_\pi^2),
\end{equation}
\begin{eqnarray}
\phi=\sqrt2\left(
\begin{array}{ccc}
\frac{\pi^0}{\sqrt2}+\frac{\eta}{\sqrt6}&\pi^+&K^+\\
\pi^-&-\frac{\pi^0}{\sqrt2}+\frac{\eta}{\sqrt6}&K^0\\
K^-&\overline{K}^0&-\frac{2}{\sqrt6}\eta
\end{array}\right),\qquad
B=\left(
\begin{array}{ccc}
\frac{\Sigma^0}{\sqrt2}+\frac{\Lambda}{\sqrt6}&\Sigma^+&p\\
\Sigma^-&-\frac{\Sigma^0}{\sqrt2}+\frac{\Lambda}{\sqrt6}&n\\
\Xi^-&\Xi^0&-\frac{2}{\sqrt6}\Lambda
\end{array}\right),
\end{eqnarray}
\begin{equation}
i{D}_\mu T^\nu_{abc}=i\partial_\mu T^\nu_{abc}+(\Gamma_\mu)^d_a
T^\nu_{dbc} + (\Gamma_\mu)^d_b T^\nu_{adc} +(\Gamma_\mu)^d_c
T^\nu_{abd},
\end{equation}
and
\begin{eqnarray}
T_{111}=\Delta^{++}, \qquad T_{112}=\frac{\Delta^{+}}{\sqrt3}, \qquad T_{122}=\frac{\Delta^{0}}{\sqrt3},
\qquad T_{222}=\Delta^{-}, \qquad T_{113}=\frac{\Sigma^{*+}}{\sqrt3},\nonumber\\
T_{123}=\frac{\Sigma^{*0}}{\sqrt6}, \qquad
T_{223}=\frac{\Sigma^{*-}}{\sqrt3}, \qquad
T_{133}=\frac{\Xi^{*0}}{\sqrt3}, \qquad
T_{233}=\frac{\Xi^{*-}}{\sqrt3}, \qquad T_{333}=\Omega^-.
\end{eqnarray}

The HB$\chi$PT Lagrangians consist of two parts at higher order
\cite{Jenkins1991}: the recoil corrections and counter terms. The
former part comes from the lower order relativistic chiral
Lagrangians. The counter terms contain LECs and are constructed
with respect to QCD symmetries. To $O(\epsilon^3)$, the following
high order Lagrangians contribute in our calculation,
\begin{eqnarray}
 {\mathcal L}_{\phi BT}^{(2)}&=&
          \frac{{\mathcal H}^2}{3M} \bar T_\mu^{ijk} v\cdot u_{ i}^{~~s} v\cdot u_{ j}^{~~l} T^\mu_{slk}
          +\frac{{\mathcal H}^2}{6M} \bar T_\mu^{ijk} v\cdot u_{ i}^{~~s} v\cdot u_{ s}^{~~l} T^\mu_{ljk}\nonumber\\
                   &&+\left\{d_1 \bar T_\mu^{ijk} v\cdot u_{ i}^{~~s} v\cdot u_{ j}^{~~l} T^\mu_{slk}
                    +d_2 \bar T_\mu^{ijk} v\cdot u_{ i}^{~~s} v\cdot u_{ s}^{~~l} T^\mu_{ljk}
                    +d_3 \Tr(v\cdot u~ v\cdot u) \bar T_\mu T^\mu
                    +d_4 \bar T_\mu \tilde \chi _+ T^\mu \right\},  \label{L2}\\
 {\mathcal L}_{\phi BT}^{(3)}&=&
        -\frac{{\mathcal H}^2}{6M^2}
                   \bar T_\mu^{ijk} v\cdot u_{ i}^{~~s} (iv\cdot D +\delta) v\cdot u_{ j}^{~~l}  T^\mu_{slk}
        -\frac{{\mathcal H}^2}{12M^2}
                   \bar T_\mu^{ijk} v\cdot u_{ i}^{~~s} (iv\cdot D +\delta) v\cdot u_{ s}^{~~l}  T^\mu_{ljk}\nonumber\\
        &&+\left\{(h_1 \bar T_\mu^{ijk} u_{\nu i}^{~~~s} u_{\rho j}^{~~~l}v^\nu i D^\rho T^\mu_{slk}+h.c.)
        +(h_2 \bar T_\mu^{ijk} u_{\nu i}^{~~~s} u_{\rho s}^{~~~l}v^\nu i D^\rho T^\mu_{ljk}+h.c.)
        +h_3 \Tr(u_\nu u_\rho) \bar T^\mu  v^\nu i D^\rho T_\mu\right.
         \nonumber \\& &\left.\quad
        +(h_4 \bar T_\mu \tilde \chi _+ iv\cdot D T^\mu+h.c.)
        +(h_5 \bar T_\mu^{ijk} \tilde \chi_{-i}^{~~~s} v\cdot u_s^{~~l} T^\mu_{ljk}+h.c.)
        \right\}, \label{L3}
\end{eqnarray}
where $\tilde \chi_\pm=\chi_\pm-\frac13 \Tr(\chi_\pm)$ and $M$ is
the mass of the nucleon. In principle, the unknown coefficients in
the Lagrangins could be determined from either QCD or the
experimental measurements. Due to the difficulty in the
non-perturbative region of QCD and the lack of experimental data,
we will resort to the resonance saturation method when estimating
LECs \cite{Ecker1989}.

\section{The $T$-matrices at thresholds} \label{sec3}

The total S-wave scattering cross section near the threshold can
be expressed in terms of the scattering length. It can also be
calculated directly using the threshold $T$-matrix $T_{PT}$. To
ensure the same total cross section in quantum mechanics, the
scattering length $a_{PT}$ is related with the threshold
$T$-matrix $T_{PT}$: $T_{PT}=4\pi(1+\frac{m_P}{M_T}) a_{PT}$,
where $m_P$ and $M_T$ are the masses of the pseudoscalar meson and
the light decuplet baryon respectively. We use a convention that
$a_{PT}>0$ ($a_{PT}<0$) if the interaction is attractive
(repulsive). There are 27 independent isospin invariant amplitudes
for the octet meson-decuplet baryon interactions. We list their
explicit expressions in the following subsections. One may get any
other amplitudes with the help of the isospin symmetry.

\begin{figure}[!htbp]
\centering
   \scalebox{0.8}{\includegraphics{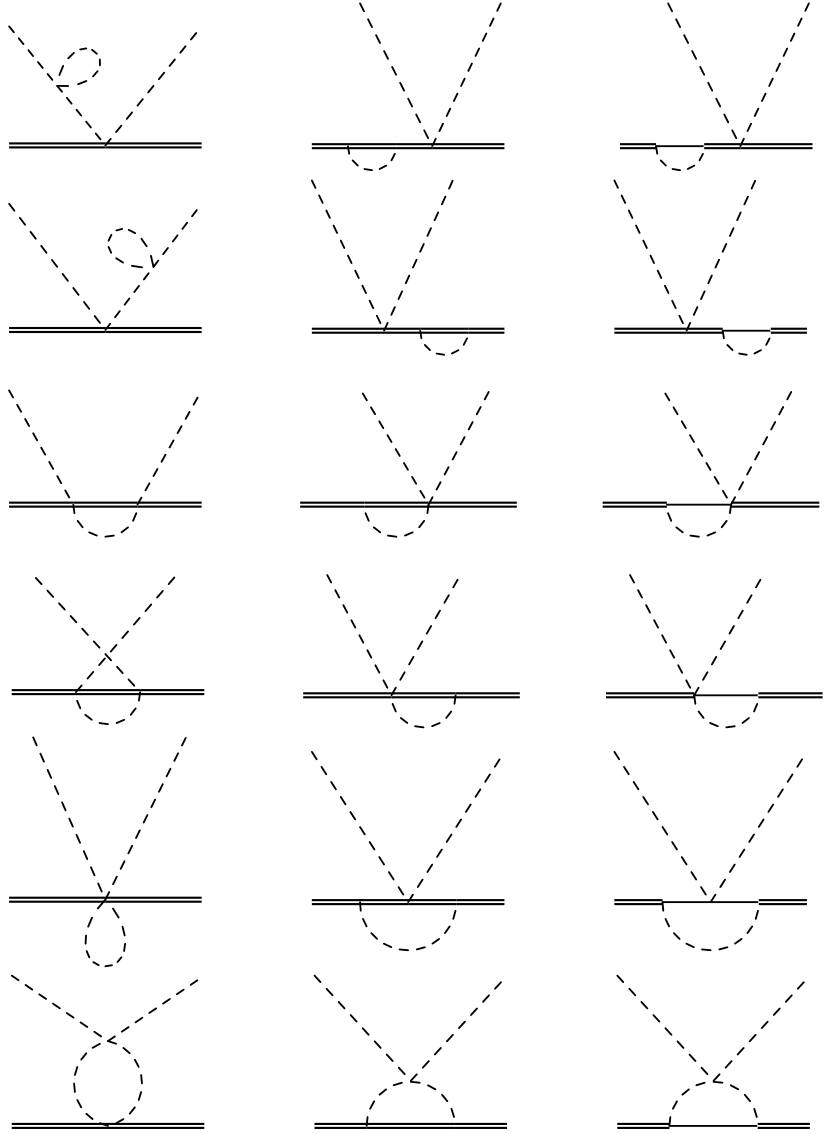}}\\
 \caption{Non-vanishing loop diagrams for the pseudoscalar meson-light decuplet baryon scattering lengths
   to $O(\epsilon^3)$ in HB$\chi$PT with explicit $\Delta$.
   Dashed lines, solid lines and double solid line represent Goldstone bosons, octet baryons and decuplet
   baryons, respectively.} \label{LoopDiag}
 \end{figure}

The first two order $T$-matrices are easy to derive from the
Lagrangians (\ref{L1}) and (\ref{L2}). At the third order, loop
corrections enter. There are 18 non-vanishing diagrams
contributing to the threshold amplitudes, which are shown in Fig.
\ref{LoopDiag}. We calculate the loop integrations with the
dimensional regularization and modified minimal subtraction. The
divergences from the loops can be absorbed through the LECs $h_i$,
\begin{equation}
  h_1={\mathcal C}^2 L +h_1^r, \quad
  h_2=\frac12 {\mathcal C}^2 L +h_2^r, \quad
  h_3=-\frac13 {\mathcal C}^2 L +h_3^r, \quad
  h_4=\frac18 {\mathcal C}^2 L +h_4^r, \quad
  h_5=-\frac94  L +h_5^r,
\end{equation}
where
\begin{equation}
  L=\frac{\lambda^{D-4}}{16\pi^2}\left\{\frac1{D-4}+\frac12(\gamma_E-1-\ln 4\pi)\right\},\quad
     (\text{Euler constant~}\gamma_E=0.5772157),
\end{equation}
and $\lambda$ is the scale from the dimensional regularization. To
make the final expressions short, we introduce a constant and two
functions,
\begin{eqnarray}
J&=&
  \frac{{\mathcal C}^2 m_K^2}{12 \pi ^2 f^4 \left(m_\eta^2-m_\pi^2\right)}
     \left\{\left( \delta ^3-\frac32 m_\eta^2 \delta \right) \log \frac{m_\eta}{\lambda}
     + (\frac32 m_\pi^2 \delta- \delta^3)\log\frac{m_\pi}{\lambda}\right.\nonumber\\
  &&\left.\qquad
     -( \delta ^2 -m_\pi^2)^{3/2} \left(\log \frac{\sqrt{\delta ^2-m_\pi^2}+\delta }{m_\pi }-i \pi\right)
     -\left(m_\eta^2-\delta ^2\right)^{3/2} \cos ^{-1}\left(-\frac{\delta }{m_\eta}\right)\right.\nonumber\\
  &&\left. \qquad
     +(m_\eta^2 - m_\pi^2) \delta    \frac{}{}
     \right\},  \label{J}\\
W(m)&=&
  \frac{{\mathcal C}^2}{16 \pi ^2 f^4}
  \begin{cases}
     -2 \sqrt{m^2-\delta ^2} \cos ^{-1}\left(-\frac{\delta }{m}\right)-2 \delta  \log\frac{m}{\lambda }+\delta
                                                                     & m>\delta  \\
     -2\delta\log\frac{m}{\lambda }
     -2\sqrt{\delta ^2-m^2}\log\frac{\sqrt{\delta ^2-m^2}+\delta }{m }
     +\delta+2 \pi i \sqrt{\delta ^2-m^2}                           & m\leq \delta
\end{cases},  \label{W}\\
V(m^2,\omega)&=&
  \frac{\omega^3 \log \frac{m}{\lambda }}{\pi ^2 f^4}-\frac{\omega^3}{2 \pi ^2 f^4}
  -\frac{ \omega^2}{\pi^2 f^4}
    \begin{cases}
      -\sqrt{m^2-\omega^2} \cos ^{-1}\left(-\frac{\omega}{m} \right)& m^2\geq \omega^2 \\
      \sqrt{\omega^2-m^2} \log \frac{\sqrt{\omega^2-m^2}-\omega}{m} & m^2<\omega^2, \omega<0 \\
      \sqrt{\omega^2-m^2} \left(-\log \frac{\sqrt{\omega^2-m^2}+\omega}{m}+i \pi \right)& m^2<\omega^2, \omega\geq 0
   \end{cases}.
\end{eqnarray}

\subsection{Pion-decuplet scattering}

There are 9 independent $T$-matrices for the pion scattering due
to the isospin symmetry. At the leading chiral order
$O(\epsilon)$, only the simplest tree diagram from the
Weinberg-Tomozawa terms is involved,
\begin{eqnarray}
  &&T^{(5/2)}_{\pi\Delta}=-\frac{3 m_\pi}{2 f_\pi^2},\quad
    T^{(3/2)}_{\pi\Delta}=\frac{m_\pi}{f_\pi^2},\quad
    T^{(1/2)}_{\pi\Delta}=\frac{5 m_\pi}{2 f_\pi^2},\quad
    T^{(2)}_{\pi\Sigma^*}=-\frac{m_\pi}{f_\pi^2},\quad
    T^{(1)}_{\pi\Sigma^*}=\frac{m_\pi}{f_\pi^2},\quad\nonumber\\
  &&T^{(0)}_{\pi\Sigma^*}=\frac{2 m_\pi}{f_\pi^2},\quad
    T^{(3/2)}_{\pi\Xi^*}=-\frac{m_\pi}{2 f_\pi^2},\quad
    T^{(1/2)}_{\pi\Xi^*}=\frac{m_\pi}{f_\pi^2},\quad
    T^{(1)}_{\pi\Omega}=0,
\end{eqnarray}
where the superscript in the bracket represents the total isospin
of the channel and $f_\pi$ is the renormalized pion decay constant
(similar to $f_K, f_\eta$ in the following text). Since we express
the $T$-matrices with $f_\pi$ rather than $f$, the difference is
of order $O(\epsilon^3)$.

The results at $O(\epsilon^2)$ also come from the tree diagram,
\begin{eqnarray}
&&
T^{(5/2)}_{\pi\Delta}=-\frac{m_\pi^2 (3 \tilde d_2+6 d_3-4 d_4)}{6 f_\pi^2},\quad
T^{(3/2)}_{\pi\Delta}=-\frac{m_\pi^2 (5 \tilde d_1+3 \tilde d_2+6 d_3-4 d_4)}{6 f_\pi^2},\quad
T^{(1/2)}_{\pi\Delta}=\frac{m_\pi^2 (4 \tilde d_1-3 \tilde d_2-6 d_3+4 d_4)}{6 f_\pi^2},\nonumber\\
  &&
T^{(2)}_{\pi\Sigma^*}=-\frac{m_\pi^2 (\tilde d_2+3 d_3)}{3 f_\pi^2},\quad
T^{(1)}_{\pi\Sigma^*}=-\frac{m_\pi^2 (\tilde d_1+\tilde d_2+3 d_3)}{3 f_\pi^2},\quad
T^{(0)}_{\pi\Sigma^*}=\frac{m_\pi^2 (3 \tilde d_1-2 \tilde d_2-6 d_3)}{6 f_\pi^2},\nonumber\\
  &&
T^{(3/2)}_{\pi\Xi^*}=-\frac{m_\pi^2 (\tilde d_2+6 d_3+4 d_4)}{6 f_\pi^2},\quad
T^{(1/2)}_{\pi\Xi^*}=-\frac{m_\pi^2 (\tilde d_2+6 d_3+4 d_4)}{6 f_\pi^2},\quad
T^{(1)}_{\pi\Omega}=-\frac{m_\pi^2 (3 d_3+4 d_4)}{3 f_\pi^2},
\label{Tpi2}
\end{eqnarray}

where
\begin{equation}
  \tilde d_1=d_1+\frac{{\mathcal H}^2}{3M}, \quad
  \tilde d_2=d_2+\frac{{\mathcal H}^2}{6M}.
\end{equation}

At $O(\epsilon^3)$, the $T$-matrices contain both loop and tree
diagram contributions. Moreover, they receive contributions from
the replacement of $f$ with $f_\pi$ in the $O(\epsilon^1)$
magnitude. Just for the simplicity of the writing and the reuse of
the expressions at $O(\epsilon^2)$, we divide $T$ into two parts
at this order,
\begin{equation}
  T=\tilde T+T_2.
  \label{TTildeT}
\end{equation}
One obtains $T_2$ from the $O(\epsilon^2)$ $T$-matrices (\ref{Tpi2}) with the following replacements
\begin{equation}
  \tilde d_1\rightarrow 2h_1^r \delta-\frac{{\mathcal H}^2 \delta}{3M^2}, \quad
  \tilde d_2\rightarrow 2h_2^r- \frac{{\mathcal H}^2 \delta}{6M^2} , \quad
  d_3\rightarrow h_3^r \delta, \quad
  d_4\rightarrow 2h_4^r \delta.
  \label{dh}
\end{equation}
$\tilde T$ are listed below,

\begin{eqnarray}
  \tilde T^{(5/2)}_{\pi\Delta}&=&-\frac{3}{16} V(m_K^2,-m_\pi)-\frac{15}{16} V(m_\pi^2,-m_\pi)-\frac{9}{16}V(m_\pi^2,m_\pi)
       -\frac{5 {\mathcal{H}}^2 m_\eta m_\pi^2}{1728 \pi  f_\pi^4}+\frac{5 {\mathcal{H}}^2 m_\pi^3}{192 \pi  f_\pi^4}
       +\frac{4 \tilde h_5 m_\pi^3}{f_\pi^2},\nonumber\\
\tilde T^{(3/2)}_{\pi\Delta}&=&-\frac{1}{32} V(m_K^2,-m_\pi)-\frac{5}{32} V(m_K^2,m_\pi)-\frac{1}{4}V(m_\pi^2,m_\pi)
      -\frac{5 {\mathcal{H}}^2 m_\eta m_\pi^2}{1728 \pi  f_\pi^4}-\frac{55 {\mathcal{H}}^2 m_\pi^3}{1728 \pi  f_\pi^4}
      -\frac{5}{6} m_\pi^2 W(m_\pi)\nonumber\\
   &&-\frac{8 \tilde h_5 m_\pi^3}{3 f_\pi^2},\nonumber\\
\tilde T^{(1/2)}_{\pi\Delta}&=&\frac{1}{16} V(m_K^2,-m_\pi)-\frac{1}{4} V(m_K^2,m_\pi)-\frac{15}{16}V(m_\pi^2,-m_\pi)
      -\frac{25}{16} V(m_\pi^2,m_\pi)-\frac{5 {\mathcal{H}}^2 m_\eta m_\pi^2}{1728 \pi  f_\pi^4}
      +\frac{125 {\mathcal{H}}^2 m_\pi^3}{1728 \pi  f_\pi^4}\nonumber\\
   && +\frac{2}{3} m_\pi^2 W(m_\pi)
      -\frac{20 \tilde h_5 m_\pi^3}{3 f_\pi^2},\nonumber\\
\tilde T^{(2)}_{\pi\Sigma^*}&=&-\frac{5}{16} V(m_K^2,-m_\pi)-\frac{3}{16} V(m_K^2,m_\pi)-\frac{1}{2}V(m_\pi^2,-m_\pi)
      -\frac{1}{4} V(m_\pi^2,m_\pi)+\frac{5 {\mathcal{H}}^2 m_\pi^3}{432 \pi  f_\pi^4}
      +\frac{1}{12} m_\pi^2 W(m_\eta)\nonumber\\
   && -\frac{1}{12} m_\pi^2 W(m_\pi)
      +\frac{8 \tilde h_5 m_\pi^3}{3 f_\pi^2},
      \nonumber\\
\tilde T^{(1)}_{\pi\Sigma^*}&=&-\frac{3}{16} V(m_K^2,-m_\pi)-\frac{5}{16} V(m_K^2,m_\pi)-\frac{1}{4}V(m_\pi^2,m_\pi)
     -\frac{5 {\mathcal{H}}^2 m_\pi^3}{432 \pi  f_\pi^4}+\frac{1}{12} m_\pi^2 W(m_\eta)-\frac{5}{12} m_\pi^2 W(m_\pi)
     \nonumber\\
  &&-\frac{8 \tilde h_5 m_\pi^3}{3 f_\pi^2}\nonumber\\
\tilde T^{(0)}_{\pi\Sigma^*}&=&-\frac{1}{8} V(m_K^2,-m_\pi)-\frac{3}{8} V(m_K^2,m_\pi)-\frac{1}{2}V(m_\pi^2,-m_\pi)
     -V(m_\pi^2,m_\pi)+\frac{5 {\mathcal{H}}^2 m_\pi^3}{108 \pi  f_\pi^4}+\frac{1}{12} m_\pi^2 W(m_\eta)\nonumber\\
  && +\frac{5}{12} m_\pi^2 W(m_\pi)
     -\frac{16 \tilde h_5 m_\pi^3}{3 f_\pi^2},\nonumber\\
\tilde T^{(3/2)}_{\pi\Xi^*}&=&-\frac{5}{16} V(m_K^2,-m_\pi)-\frac{1}{4} V(m_K^2,m_\pi)-\frac{3}{16}V(m_\pi^2,-m_\pi)
     -\frac{1}{16} V(m_\pi^2,m_\pi)-\frac{5 {\mathcal{H}}^2 m_\eta m_\pi^2}{1728 \pi  f_\pi^4}
     +\frac{5 {\mathcal{H}}^2 m_\pi^3}{1728 \pi  f_\pi^4}\nonumber\\
  && +\frac{1}{12} m_\pi^2 W(m_\eta)
     -\frac{1}{12} m_\pi^2 W(m_\pi)
     +\frac{4 \tilde h_5 m_\pi^3}{3 f_\pi^2},\nonumber\\
\tilde T^{(1/2)}_{\pi\Xi^*}&=&-\frac{7}{32} V(m_K^2,-m_\pi)-\frac{11}{32} V(m_K^2,m_\pi)-\frac{1}{4}V(m_\pi^2,m_\pi)
     -\frac{5 {\mathcal{H}}^2 m_\eta m_\pi^2}{1728 \pi  f_\pi^4}+\frac{5 {\mathcal{H}}^2 m_\pi^3}{1728 \pi  f_\pi^4}
     +\frac{1}{12} m_\pi^2 W(m_\eta)\nonumber\\
  && -\frac{1}{12} m_\pi^2 W(m_\pi)
     -\frac{8 \tilde h_5 m_\pi^3}{3 f_\pi^2}\nonumber\\
\tilde T^{(1)}_{\pi\Omega}&=&-\frac{3}{16} V(m_K^2,-m_\pi)-\frac{3}{16} V(m_K^2,m_\pi)
     -\frac{5 {\mathcal{H}}^2 m_\eta m_\pi^2}{432 \pi  f_\pi^4},
\end{eqnarray}
where
\begin{equation}
  \tilde h_5=h_5^r-\frac{{\mathcal H}^2}{96 M^2}.
\end{equation}

\subsection{Kaon-decuplet scattering}

At the leading order, there are 14 independent $T$-matrices, which are
\begin{eqnarray}
&&T^{(2)}_{K\Delta}=-\frac{3 m_K}{2 f_K^2},\quad
T^{(1)}_{K\Delta}=\frac{m_K}{2 f_K^2},\quad
T^{(3/2)}_{K\Sigma^*}=-\frac{m_K}{2 f_K^2},\quad
T^{(1/2)}_{K\Sigma^*}=\frac{m_K}{f_K^2},\quad
T^{(1)}_{K\Xi^*}=\frac{m_K}{2 f_K^2},\quad \nonumber\\
&&T^{(0)}_{K\Xi^*}=\frac{3 m_K}{2 f_K^2},\quad
T^{(1/2)}_{K\Omega}=\frac{3 m_K}{2 f_K^2},\quad
T^{(2)}_{\bar K\Delta}=0,\quad
T^{(1)}_{\bar K\Delta}=\frac{2 m_K}{f_K^2},\quad
T^{(3/2)}_{\bar K\Sigma^*}=-\frac{m_K}{2 f_K^2},\quad\nonumber\\
&&T^{(1/2)}_{\bar K\Sigma^*}=\frac{m_K}{f_K^2},\quad
T^{(1)}_{\bar K\Xi^*}=-\frac{m_K}{f_K^2},\quad
T^{(0)}_{\bar K\Xi^*}=0,\quad
T^{(1/2)}_{\bar K\Omega}=-\frac{3 m_K}{2 f_K^2}.
\end{eqnarray}

At the next leading order, we have
\begin{eqnarray}
&&
T^{(2)}_{K\Delta}=-\frac{m_K^2 (3 \tilde d_2+6 d_3-4 d_4)}{6 f_K^2},\quad
T^{(1)}_{K\Delta}=\frac{m_K^2 (\tilde d_2-6 d_3-12 d_4)}{6 f_K^2},\quad
T^{(3/2)}_{K\Sigma^*}=-\frac{m_K^2 (2 \tilde d_1+3 \tilde d_2+6 d_3-4 d_4)}{6 f_K^2},\nonumber\\
  &&
T^{(1/2)}_{K\Sigma^*}=\frac{m_K^2 (\tilde d_1-6 d_3-8 d_4)}{6 f_K^2},\quad
T^{(1)}_{K\Xi^*}=-\frac{m_K^2 (2 \tilde d_1+3 \tilde d_2+6 d_3-4 d_4)}{6 f_K^2},\quad
T^{(0)}_{K\Xi^*}=\frac{m_K^2 (2 \tilde d_1-\tilde d_2-6 d_3-4 d_4)}{6 f_K^2},\nonumber\\
&&
T^{(1/2)}_{K\Omega}=-\frac{m_K^2 (3 \tilde d_2+6 d_3-4 d_4)}{6 f_K^2},\quad
T^{(2)}_{\bar K\Delta}=-\frac{m_K^2 (3 d_3+4 d_4)}{3 f_K^2},\quad
T^{(1)}_{\bar K\Delta}=-\frac{m_K^2 (2 \tilde d_2+3 d_3-4 d_4)}{3 f_K^2},\nonumber\\
&&
T^{(3/2)}_{\bar K\Sigma^*}=-\frac{m_K^2 (\tilde d_2+6 d_3+4 d_4)}{6 f_K^2},\quad
T^{(1/2)}_{\bar K\Sigma^*}=-\frac{m_K^2 (3 \tilde d_1+4 \tilde d_2+6 d_3-8 d_4)}{6 f_K^2},\quad
T^{(1)}_{\bar K\Xi^*}=-\frac{m_K^2 (\tilde d_2+3 d_3)}{3 f_K^2},\nonumber\\
&&
T^{(0)}_{\bar K\Xi^*}=-\frac{m_K^2 (2 \tilde d_1+2 \tilde d_2+3 d_3-4 d_4)}{3 f_K^2},\quad
T^{(1/2)}_{\bar K\Omega}=-\frac{m_K^2 (3 \tilde d_2+6 d_3-4 d_4)}{6 f_K^2}.
\end{eqnarray}

At the next-next leading order, we just list $\tilde T$ here. One
may get the total $T$-matrices from Eqs. (\ref{TTildeT}) and
(\ref{dh}).
\begin{eqnarray}
\tilde T^{(2)}_{K\Delta}&=&-\frac{3}{4} V(m_K^2,-m_K)-\frac{9}{16} V(m_K^2,m_K)-\frac{9}{32} V(m_\eta^2,-m_K)
     -\frac{3}{32} V(m_\pi^2,-m_K)+\frac{5 {\mathcal{H}}^2 m_K^2 m_\pi^2}{432 \pi  f_K^4 (m_\eta+m_\pi)}\nonumber\\&&
     +\frac{5 {\mathcal{H}}^2 m_K^2 m_\eta}{288 \pi  f_K^4}
     +\frac{4 \tilde h_5 m_K^3}{f_K^2},\nonumber\\
\tilde T^{(1)}_{K\Delta}&=&\frac{1}{4} V(m_K^2,-m_K)-\frac{1}{16} V(m_K^2,m_K)+\frac{3}{32} V(m_\eta^2,-m_K)
     -\frac{7}{32} V(m_\pi^2,-m_K)-\frac{25 {\mathcal{H}}^2 m_K^2 m_\pi^2}{1296 \pi  f_K^4 (m_\eta+m_\pi)}\nonumber\\&&
     -\frac{35 {\mathcal{H}}^2 m_K^2 m_\eta}{2592 \pi  f_K^4}
     -\frac{4 \tilde h_5 m_K^3}{3 f_K^2},\nonumber\\
\tilde T^{(3/2)}_{K\Sigma^*}&=&-\frac{3}{16} V(m_K^2,-m_K)-\frac{1}{16} V(m_K^2,m_K)-\frac{3}{8} V(m_\eta^2,-m_K)
     -\frac{9}{32} V(m_\eta^2,m_K)-\frac{1}{8} V(m_\pi^2,-m_K)\nonumber\\&&
     -\frac{5}{32} V(m_\pi^2,m_K)
     -\frac{J}{6}
     -\frac{1}{6} m_K^2 W(m_\eta)
     +\frac{4 \tilde h_5 m_K^3}{3 f_K^2},
     \nonumber\\
\tilde T^{(1/2)}_{K\Sigma^*}&=&-\frac{1}{4} V(m_K^2,m_K)+\frac{3}{16} V(m_\eta^2,-m_K)-\frac{5}{16} V(m_\pi^2,-m_K)
     -\frac{1}{4} V(m_\pi^2,m_K)+\frac{J}{3}-\frac{1}{6} m_K^2 W(m_\eta)
     -\frac{8 \tilde h_5 m_K^3}{3 f_K^2},\nonumber\\
\tilde T^{(1)}_{K\Xi^*}&=&-\frac{1}{8} V(m_K^2,-m_K)-\frac{1}{16} V(m_K^2,m_K)-\frac{9}{32} V(m_\eta^2,-m_K)
     -\frac{3}{8} V(m_\eta^2,m_K)-\frac{3}{32} V(m_\pi^2,-m_K)-\frac{1}{4} V(m_\pi^2,m_K)\nonumber\\&&
     -\frac{5 {\mathcal{H}}^2 m_K^2 m_\pi^2}{1296 \pi  f_K^4 (m_\eta+m_\pi)}
     +\frac{5 {\mathcal{H}}^2 m_K^2 m_\eta}{2592 \pi  f_K^4}-\frac{J}{6}
     -\frac{1}{6} m_K^2 W(m_\eta)
     -\frac{4 \tilde h_5 m_K^3}{3 f_K^2},\nonumber\\
\tilde T^{(0)}_{K\Xi^*}&=&-\frac{3}{8} V(m_K^2,-m_K)-\frac{9}{16} V(m_K^2,m_K)+\frac{9}{32} V(m_\eta^2,-m_K)
     -\frac{9}{32} V(m_\pi^2,-m_K)-\frac{3}{8} V(m_\pi^2,m_K)\nonumber\\&&
     +\frac{5 {\mathcal{H}}^2 m_K^2 m_\pi^2}{432 \pi  f_K^4 (m_\eta+m_\pi)}
     +\frac{5 {\mathcal{H}}^2 m_K^2 m_\eta}{288 \pi  f_K^4}+\frac{J}{2}
     -\frac{1}{6} m_K^2 W(m_\eta)
     -\frac{4 \tilde h_5 m_K^3}{f_K^2},\nonumber\\
\tilde T^{(1/2)}_{K\Omega}&=&-\frac{9}{16} V(m_K^2,-m_K)-\frac{9}{16} V(m_K^2,m_K)-\frac{9}{32} V(m_\eta^2,m_K)
     -\frac{9}{32} V(m_\pi^2,m_K)+\frac{5 {\mathcal{H}}^2 m_K^2 m_\eta}{216 \pi  f_K^4}
     -\frac{4 \tilde h_5 m_K^3}{f_K^2},\nonumber\\
\tilde T^{(2)}_{\bar K\Delta}&=&-\frac{3}{16} V(m_K^2,-m_K)-\frac{3}{16} V(m_\pi^2,m_K)
      -\frac{5 {\mathcal{H}}^2 m_K^2 m_\pi^2}{432 \pi  f_K^4 (m_\eta+m_\pi)}
      -\frac{5 {\mathcal{H}}^2 m_K^2 m_\eta}{864 \pi  f_K^4},\nonumber\\
\tilde T^{(1)}_{\bar K\Delta}&=&-\frac{11}{16} V(m_K^2,-m_K)-V(m_K^2,m_K)-\frac{3}{8} V(m_\eta^2,m_K)
      -\frac{1}{16} V(m_\pi^2,m_K)+\frac{25 {\mathcal{H}}^2 m_K^2 m_\pi^2}{1296 \pi  f_K^4 (m_\eta+m_\pi)}\nonumber\\&&
      +\frac{65 {\mathcal{H}}^2 m_K^2 m_\eta}{2592 \pi  f_K^4}
      -\frac{16 \tilde h_5 m_K^3}{3 f_K^2},\nonumber\\
\tilde T^{(3/2)}_{\bar K\Sigma^*}&=&-\frac{3}{16} V(m_K^2,-m_K)-\frac{1}{16} V(m_K^2,m_K)-\frac{3}{32} V(m_\eta^2,-m_K)
     -\frac{7}{32} V(m_\pi^2,-m_K)-\frac{1}{4} V(m_\pi^2,m_K)+\frac{J}{6}\nonumber\\&&
     -\frac{1}{6} m_K^2 W(m_\eta)
     +\frac{4 \tilde h_5 m_K^3}{3 f_K^2},\nonumber\\
\tilde T^{(1/2)}_{\bar K\Sigma^*}&=&-\frac{1}{4} V(m_K^2,m_K)-\frac{3}{8} V(m_\eta^2,-m_K)-\frac{9}{16} V(m_\eta^2,m_K)
     -\frac{1}{8} V(m_\pi^2,-m_K)-\frac{1}{16} V(m_\pi^2,m_K)-\frac{J}{3}\nonumber\\&&
     -\frac{1}{6} m_K^2 W(m_\eta)
     -\frac{8 \tilde h_5 m_K^3}{3 f_K^2},\nonumber\\
\tilde T^{(1)}_{\bar K\Xi^*}&=&-\frac{5}{16} V(m_K^2,-m_K)-\frac{1}{4} V(m_K^2,m_K)-\frac{3}{16} V(m_\eta^2,-m_K)
     -\frac{5}{16} V(m_\pi^2,-m_K)-\frac{3}{16} V(m_\pi^2,m_K)\nonumber\\&&
     +\frac{5 {\mathcal{H}}^2 m_K^2 m_\pi^2}{1296 \pi  f_K^4 (m_\eta+m_\pi)}
     +\frac{25 {\mathcal{H}}^2 m_K^2 m_\eta}{2592 \pi  f_K^4}+\frac{J}{6}-\frac{1}{6} m_K^2 W(m_\eta)
     +\frac{8 \tilde h_5 m_K^3}{3 f_K^2},\nonumber\\
\tilde T^{(0)}_{\bar K\Xi^*}&=&\frac{3}{16} V(m_K^2,-m_K)-\frac{9}{16} V(m_\eta^2,-m_K)-\frac{9}{16} V(m_\eta^2,m_K)
    -\frac{3}{16} V(m_\pi^2,-m_K)-\frac{5 {\mathcal{H}}^2 m_K^2 m_\pi^2}{432 \pi  f_K^4 (m_\eta+m_\pi)}\nonumber\\&&
    -\frac{5 {\mathcal{H}}^2 m_K^2 m_\eta}{864 \pi  f_K^4}-\frac{J}{2}
    -\frac{1}{6} m_K^2 W(m_\eta),\nonumber\\
\tilde T^{(1/2)}_{\bar K\Omega}&=&-\frac{9}{16} V(m_K^2,-m_K)-\frac{9}{16} V(m_K^2,m_K)-\frac{9}{32} V(m_\eta^2,-m_K)
   -\frac{9}{32} V(m_\pi^2,-m_K)+\frac{5 {\mathcal{H}}^2 m_K^2 m_\eta}{216 \pi  f_K^4}\nonumber\\&&
   +\frac{4 \tilde h_5 m_K^3}{f_K^2}.
\end{eqnarray}

\subsection{Eta-decuplet scattering}

Eta and decuplet-baryon $T$-matrices start at the second chiral order,
\begin{eqnarray}
&&
T^{(3/2)}_{\eta\Delta}=-\frac{(\tilde d_1+\tilde d_2 +6 d_3+8d_4)m_\eta^2}{6 f_\eta^2}
            +\frac{2 d_4 m_\pi^2}{3 f_\eta^2},\quad
T^{(1)}_{\eta\Sigma^*}=\frac{(\tilde d_1-2\tilde d_2-6 d_3) m_\eta^2}{6 f_\eta^2},\nonumber\\&&
T^{(1/2)}_{\eta\Xi^*}=-\frac{(3\tilde d_2+6d_3-8 d_4) m_\eta^2}{6 f_\eta^2}
     -\frac{2 d_4 m_\pi^2}{3 f_\eta^2},\quad
T^{(0)}_{\eta\Omega}=-\frac{(2 \tilde d_1+2 \tilde d_2+ 3d_3-8 d_4) m_\eta^2}{3 f_\eta^2}
     -\frac{4 d_4 m_\pi^2}{3 f_\eta^2}.
\end{eqnarray}

At the third order, we have
\begin{eqnarray}
\tilde T^{(3/2)}_{\eta\Delta}&=&-\frac{9}{32} V(m_K^2,-m_\eta)-\frac{9}{32} V(m_K^2,m_\eta)
    +\frac{5 {\mathcal{H}}^2 m_K^3}{432 \pi  f_\eta^4}-\frac{5 {\mathcal{H}}^2 m_K^2 m_\eta}{324 \pi  f_\eta^4}
    +\frac{35 {\mathcal{H}}^2 m_\eta m_\pi^2}{5184 \pi  f_\eta^4}-\frac{25 {\mathcal{H}}^2 m_\pi^3}{1728 \pi  f_\eta^4}
    -\frac{1}{3} m_K^2 W(m_K)\nonumber\\&&
    +\frac{1}{6} m_\pi^2 W(m_\pi),\nonumber\\
\tilde T^{(1)}_{\eta\Sigma^*}&=&-\frac{3}{4} V(m_K^2,-m_\eta)-\frac{3}{4} V(m_K^2,m_\eta)
    +\frac{5 {\mathcal{H}}^2 m_K^3}{162 \pi  f_\eta^4}-\frac{5 {\mathcal{H}}^2 m_\pi^3}{648 \pi  f_\eta^4}
    +\frac{4}{9} m_K^2 W(m_\eta)-\frac{2}{9} m_K^2 W(m_K)-\frac{7}{36} m_\pi^2 W(m_\eta)\nonumber\\&&
    +\frac{5}{36} m_\pi^2 W(m_\pi),\nonumber\\
\tilde T^{(1/2)}_{\eta\Xi^*}&=&-\frac{27}{32} V(m_K^2,-m_\eta)-\frac{27}{32} V(m_K^2,m_\eta)
    +\frac{5 {\mathcal{H}}^2 m_K^3}{144 \pi  f_\eta^4}-\frac{5 {\mathcal{H}}^2 m_K^2 m_\eta}{324 \pi  f_\eta^4}
    +\frac{35 {\mathcal{H}}^2 m_\eta m_\pi^2}{5184 \pi  f_\eta^4}-\frac{5 {\mathcal{H}}^2 m_\pi^3}{1728 \pi  f_\eta^4}
    +\frac{4}{9} m_K^2 W(m_\eta)\nonumber\\&&
    -\frac{1}{3} m_K^2 W(m_K)-\frac{7}{36} m_\pi^2 W(m_\eta)
    +\frac{1}{12} m_\pi^2 W(m_\pi),\nonumber\\
\tilde T^{(0)}_{\eta\Omega}&=&-\frac{9}{16} V(m_K^2,-m_\eta)-\frac{9}{16} V(m_K^2,m_\eta)
    +\frac{5 {\mathcal{H}}^2 m_K^3}{216 \pi  f_\eta^4}-\frac{5 {\mathcal{H}}^2 m_K^2 m_\eta}{81 \pi  f_\eta^4}
    +\frac{35 {\mathcal{H}}^2 m_\eta m_\pi^2}{1296 \pi  f_\eta^4}-\frac{2}{3} m_K^2 W(m_K).
\end{eqnarray}

We have checked that our results satisfy the $SU(3)$ symmetry and
the crossing symmetry. Moreover, we have calculated the $T$-matrices
of all 80 channels explicitly.

\section{Low-energy constants}\label{sec4}

Before the numerical calculation, we have to determine the
coupling constants in the Lagrangian. For those in the leading
order, we use \cite{PDG2010,Jenkins1991}
\begin{eqnarray}
 && m_\pi=139.57~{\rm MeV}, \quad m_K=493.68~{\rm MeV}, \quad m_\eta=547.75~{\rm MeV},
        \quad \delta=294~{\rm MeV},\nonumber\\
 &&f_\pi=92.4~{\rm MeV},\quad f_K=113~{\rm MeV},\quad f_\eta=1.2 f_K, \nonumber\\
 &&{\mathcal C}=-1.5,\quad {\mathcal H}=-1.9.
\end{eqnarray}

At the second chiral order, $d_4$ could be determined from the mass splitting of the light decuplet,
\begin{equation}
  d_4=\frac{3(m_\Delta-m_{\Sigma^*})}{4(m_\pi^2-m_K^2)}=0.51~{\rm GeV^{-1}}.  \label{d4}
\end{equation}
Other LECs can not be determined from the available experimental
data. Thus we try to estimate them with the resonance saturation
method, which was originally used in the meson case
\cite{Ecker1989} and applied to the pion nucleon interaction
latter \cite{Bernard1993}. According to this method, the effects
from higher resonances are encoded in the LECs. Here we consider
contributions from baryon resonances below 1.6 GeV and ignore
contributions from higher baryons.

The first baryon multiplet contributing to LECs is the $J^P=3/2^-$
octet, which contains $N(1520)$, $\Lambda(1690)$, $\Sigma(1670)$,
and $\Xi(1820)$. We simply denote it as $N_{1520}^\mu$ here. The
relevant interacting Lagrangian is
\begin{equation}
  {\cal L}_{N_{1520}}=
                      i G_{N(1520)} ( \bar N_{1520}^\mu \sla u T_\mu -\bar T_\mu \sla u N_{1520}^\mu).
  \label{LN1520}
\end{equation}
For the complete relativistic Lagrangian, additional terms may be constructed according to the point
transformation \cite{Tang1996}. Because we are considering the case at threshold and the external baryons are on
shell, such terms do not contribute and we ignore them here. By integrating out $N_{1520}$, one gets
\begin{equation}
  {\cal L}_{\text{eff}}^{N(1520)}\sim \frac{G_{N(1520)}^2}{m_{1520}-m_\Delta}
                        (\bar T_\mu^{ijk} v\cdot u_{ i}^{~~s} v\cdot u_{ j}^{~~l} T^\mu_{slk}
                        -\bar T_\mu^{ijk} v\cdot u_{ i}^{~~s} v\cdot u_{ s}^{~~l} T^\mu_{ljk}).
   \label{Leff1520}
\end{equation}

The $J^P=3/2^+$ decuplet containing $\Delta(1600)$ is the lightest one contributing to LECs, and is denoted as
$T_{1600}$ here. The interacting Lagrangian reads
\begin{equation}
  {\cal L}_{T_{1600}}= G_{\Delta(1600)}(\bar T_{1600}^\mu \sla u \gamma_5 T_\mu
                                  +\bar T^\mu \sla u \gamma_5 T_{1600\mu}).
  \label{LT1600}
\end{equation}
The contribution to LECs may be derived from the resultant
effective term,
 \begin{equation}
   {\cal L}_{\text{eff}}^{\Delta(1600)}\sim \frac{G_{\Delta(1600)}^2}{m_{1600}+m_\Delta}
                                     (\frac23 \bar T_\mu^{ijk} v\cdot u_{ i}^{~~s} v\cdot u_{ j}^{~~l} T^\mu_{slk}
                        +\frac13 \bar T_\mu^{ijk} v\cdot u_{ i}^{~~s} v\cdot u_{ s}^{~~l} T^\mu_{ljk}).
   \label{Leff1600}
 \end{equation}

Recall the fact that the time component of the on-shell
Rarita-Schwinger field vanishes in the static limit, there is no
chiral coupling with $J=1/2$ baryon and pion at threshold. For the
$\pi N$ interaction \cite{Bernard1993}, both the intermediate
$\Delta$ and $N^*$ baryons contribute to LECs because they are
off-shell. In the present case, the external $J=3/2$ baryon is
on-shell and static and the intermediate excited $J=1/2$ baryons
do not contribute. As a result, we only need to consider the
intermediate $J=3/2$ baryons.

Besides the baryon resonances, the scalar meson $\sigma$ would
also contribute to LECs through t-channel. From the Lagrangians,
\begin{eqnarray}
  {\cal L}_{\sigma \pi\pi}&=&\tilde c_d \Tr(u\cdot u) \sigma+\tilde c_m \Tr(\chi_+) \sigma , \nonumber\\
  {\cal L}_{\sigma \Delta\Delta} &=& g_{\sigma} \bar T_\mu T^\mu \sigma,
\end{eqnarray}
we get
\begin{equation}
  {\cal L}_{\text{eff}}^{\sigma}\sim (\frac{2g_{\sigma} \tilde c_d}{m_{600}^2}-  \frac{8g_{\sigma} \tilde c_m}{m_{600}^2})\Tr(v\cdot u~ v\cdot u) \bar T_\mu T^\mu. \label{Leff600}
\end{equation}
Similarly, the scalar octet containing $\kappa(800)$, $a_0(980)$
and $f_0(980)$ may also be considered. We denote it as $\kappa$.
The corresponding Lagrangians are
\begin{eqnarray}
  {\cal L}_{\kappa \pi\pi}&=&c_d \Tr(u\cdot u \kappa)+c_m \Tr(\chi_+ \kappa), \nonumber\\
  {\cal L}_{\kappa \Delta\Delta} &=& g_{\kappa} \bar T_\mu \kappa T^\mu,
\end{eqnarray}
from which one obtains
\begin{equation}
  {\cal L}_{\text{eff}}^{\kappa}\sim \frac{2g_{\kappa} c_d}{m_{800}^2}
                                      \bar T_\mu^{ijk} v\cdot u_{ i}^{~~s} v\cdot u_{ s}^{~~l} T^\mu_{ljk}
                      - \frac{2g_{\kappa} c_d}{3m_{800}^2} \Tr(v\cdot u~ v\cdot u) \bar T_\mu T^\mu
                      +\frac{2g_{\kappa} c_m}{m_{800}^2}  \bar T_\mu \tilde \chi_+ T^\mu.
                      \label{Leff800}
\end{equation}
In the pion-nucleon case, the vector meson $\rho$ does not contribute because the
contraction of the $\rho$-meson propagator with the corresponding
$\rho\pi\pi$ matrix element vanishes in the forward direction \cite{Bernard1993}. Here the vector mesons do not contribute because of the same reason.

Adding the above effective Lagrangians (\ref{Leff1520}), (\ref{Leff1600}), (\ref{Leff600}), and (\ref{Leff800})
together, we have
\begin{eqnarray}
  {\cal L}_{\text{eff}}&=&
                  \left(\frac{G_{N(1520)}^2}{m_{1520}-m_\Delta}+\frac{2G_{\Delta(1600)}^2}{3(m_{1600}+m_\Delta)}\right)
                             \bar T_\mu^{ijk} v\cdot u_{ i}^{~~s} v\cdot u_{ j}^{~~l} T^\mu_{slk}\nonumber\\
                        &&+\left(\frac{G_{\Delta(1600)}^2}{3(m_{1600}+m_\Delta)}
                               -\frac{G_{N(1520)}^2}{m_{1520}-m_\Delta}
                               +\frac{2g_{\kappa} c_d}{m_{800}^2}\right)
                                \bar T_\mu^{ijk} v\cdot u_{ i}^{~~s} v\cdot u_{ s}^{~~l} T^\mu_{ljk} \nonumber\\
                        &&+\left( \frac{2g_{\sigma} \tilde c_d}{m_{600}^2}
                            -  \frac{8g_{\sigma} \tilde c_m}{m_{600}^2}- \frac{2g_{\kappa} c_d}{3m_{800}^2}\right)
                                 \Tr(v\cdot u~ v\cdot u) \bar T_\mu T^\mu
                          + \frac{2g_{\kappa} c_m}{m_{800}^2}  \bar T_\mu \tilde \chi_+ T^\mu.
\end{eqnarray}
Now one estimates the LECs by comparing it with the Lagrangian
(\ref{L2}),
\begin{eqnarray}
  &&d_1=\frac{G_{N(1520)}^2}{m_{1520}-m_\Delta}+\frac{2G_{\Delta(1600)}^2}{3(m_{1600}+m_\Delta)}, \quad
  d_2=\frac{G_{\Delta(1600)}^2}{3(m_{1600}+m_\Delta)}-\frac{G_{N(1520)}^2}{m_{1520}-m_\Delta}
      +\frac{2g_{\kappa} c_d}{m_{800}^2},  \nonumber\\
  &&d_3= \frac{2g_{\sigma} \tilde c_d}{m_{600}^2}
       -  \frac{8g_{\sigma} \tilde c_m}{m_{600}^2}- \frac{2g_{\kappa} c_d}{3m_{800}^2}, \quad
  d_4=\frac{2g_{\kappa} c_m}{m_{800}^2}. \label{lec2}
\end{eqnarray}

It is not difficult to determine the coupling constant $G_{N(1520)}$ according to the decay width. From the
Lagrangian (\ref{LN1520}) and the decay to $\pi\Delta$, one deduces
\begin{equation}
  G_{N(1520)}^2=\frac{36 \pi \Gamma(N(1520)\rightarrow \Delta\pi) m_{1520}m_\Delta^2 f_\pi^2}
                      {\left|\vec p_d\right| (m_{1520}-m_\Delta)^2 (E_\Delta  +  m_\Delta)
                      (2 E_\Delta ^2 +2 E_\Delta  m_\Delta + 5m_\Delta ^2)  }.
 \label{G1520}
\end{equation}
Here $\vec p_d$ is the decay momentum, $E_\pi=\sqrt{\vec
p_d^{~2}+m_\pi^2}$, and $E_\Delta=\sqrt{\vec
p_d^{~2}+m_\Delta^2}$. From the PDG \cite{PDG2010}, one derives
\begin{equation}
  \frac{G_{N(1520)}^2 }{m_{1520}-m_\Delta}=0.27~{\rm GeV^{-1}}.
  \label{GN1520f}
\end{equation}
With the same procedure, we derive the coupling constant
$G_{\Delta(1600)}$,
\begin{equation}
  \frac{G_{\Delta(1600)}^2 }{m_{1600}+m_\Delta}=0.46~{\rm GeV^{-1}}.
  \label{GT1600f}
\end{equation}

For the coupling constants $c_d$ and $c_m$, we use \cite{Ecker1989}
\begin{equation}
  \left|c_d\right|=3.2\times10^{-2}~{\rm GeV}, \quad
  \left|c_m\right|=4.2\times10^{-2}~{\rm GeV}, \quad
  c_d c_m>0.
\end{equation}
Although there is no empirical value of $g_\kappa$, one may estimate it by comparing $d_4$ obtained with the
resonance saturation method in (\ref{lec2}) with that from the mass splitting in (\ref{d4}),
\begin{equation}
  \left|g_\kappa \right|=3.9, \qquad g_\kappa c_m>0.
\end{equation}
In addition, the coupling constants should obey the nonet relations in the large $N_c$ limit,
\begin{equation}
  \tilde c_d=\frac{\zeta}{\sqrt 3}c_d, \quad
  \tilde c_m=\frac{\zeta}{\sqrt 3}c_m, \quad
  g_\sigma=\frac{\zeta}{\sqrt 3}g_\kappa, \quad
  \zeta=\pm 1.
\end{equation}

By combining the above relations, the estimated LECs at
$O(\epsilon^2)$ are
\begin{equation}
  d_1=0.58~{\rm GeV^{-1}},\quad
  d_2=0.28~{\rm GeV^{-1}},\quad
  d_3=-1.11~{\rm GeV^{-1}},\quad
  d_4=0.51~{\rm GeV^{-1}}.
\end{equation}

In a similar way, one may consider the high order corrections and
estimate the ${\cal O}(\epsilon^3)$ LECs $h_1$--$h_5$. Due to the
uncertainty of this method, we tend to neglect these counter terms
in the following calculation. This assumption was adopted in the
study of the meson-baryon scattering lengths in Refs.
\cite{Kaiser2001,Liu2007}. The counter terms of the $\pi N$
scattering lengths are found to be much smaller than the loop
contributions at this order in Ref. \cite{Bernard1993}. However,
to explore the orders of the counter terms, we have made a very
crude estimate about these LECs $h_1$--$h_5$, which are collected
in the Appendix \ref{append}.

\section{Numerical results and discussions}\label{sec5}

We set $\lambda$ at $4\pi f_\pi$, $4\pi f_K$ and $4\pi f_\eta$
respectively for the pion, kaon and $\eta$-scattering in our
numerical analysis. The results for the $T$-matrices and the
scattering lengths at different orders are shown in Table
\ref{piT}, \ref{KT} and \ref{etaT}. The positive sign of
$a^{(3/2)}_{\pi\Delta}$, $a^{(1/2)}_{\pi\Delta}$,
$a^{(1)}_{\pi\Sigma^*}$, $a^{(0)}_{\pi\Sigma^*}$,
$a^{(1/2)}_{\pi\Xi^*}$, $a^{(1)}_{K\Delta}$,
$a^{(1/2)}_{K\Sigma^*}$,  $a^{(1)}_{K\Xi^*}$, $a^{(0)}_{K\Xi^*}$,
$a^{(1/2)}_{K\Omega}$, $a^{(1)}_{\bar K\Delta}$,
$a^{(1/2)}_{\bar K\Sigma^*}$,
$a^{(1)}_{\eta \Sigma^*}$ and
$a^{(1/2)}_{\eta \Xi^*}$ indicates that the strong
interactions for these channels are attractive.

\begin{table}
\caption{Pion-light decuplet baryon threshold $T$-matrices order by order in unit of fm. }\label{piT}
\begin{tabular}{cccccc}
\hline
                                      &$O(\epsilon^1)$          &$O(\epsilon^2)$         &$O(\epsilon^3)$          &Total             &Scattering Length  \\\hline
$T^{(5/2)}_{\pi\Delta}$               &-4.84                    &0.445                   &-1.37                    &-5.76             &-0.412             \\
$T^{(3/2)}_{\pi\Delta}$               &3.23                     &-0.253                  &-0.901-1.02$i$           &2.07-1.02$i$      &0.148-0.0728$i$    \\
$T^{(1/2)}_{\pi\Delta}$               &8.06                     &1.                      &1.44+0.814$i$            &10.5+0.814$i$     &0.751+0.0582$i$    \\
$T^{(2)}_{\pi\Sigma^*}$               &-3.23                    &0.361                   &-2.52-0.102$i$           &-5.38-0.102$i$    &-0.389-0.00736$i$  \\
$T^{(1)}_{\pi\Sigma^*}$               &3.23                     &0.0819                  &-1.8-0.509$i$            &1.51-0.509$i$     &0.109-0.0368$i$    \\
$T^{(0)}_{\pi\Sigma^*}$               &6.45                     &0.781                   &-0.412+0.509$i$          &6.82+0.509$i$     &0.493+0.0368$i$    \\
$T^{(3/2)}_{\pi\Xi^*}$                &-1.61                    &0.277                   &-2.69-0.102$i$           &-4.02-0.102$i$    &-0.293-0.00742$i$  \\
$T^{(1/2)}_{\pi\Xi^*}$                &3.23                     &0.277                   &-1.89-0.102$i$           &1.61-0.102$i$     &0.118-0.00742$i$   \\
$T^{(1)}_{\pi\Omega}$                 &0                        &0.194                   &-1.87                    &-1.68             &-0.123             \\
\hline
\end{tabular}
\end{table}

\begin{table}
\caption{Kaon-light decuplet baryon threshold $T$-matrices order by order in unit of fm. }\label{KT}
\begin{tabular}{cccccc}
\hline
                                      &$O(\epsilon^1)$          &$O(\epsilon^2)$         &$O(\epsilon^3)$          &Total             &Scattering Length  \\\hline
$T^{(2)}_{K\Delta}$                   &-11.4                    &3.73                    &-9.27                    &-17.              &-0.965             \\
$T^{(1)}_{K\Delta}$                   &3.81                     &0.917                   &3.67                     &8.4               &0.477              \\
$T^{(3/2)}_{K\Sigma^*}$               &-3.81                    &1.39                    &-9.44+6.77$i$            &-11.9+6.77$i$     &-0.696+0.397$i$    \\
$T^{(1/2)}_{K\Sigma^*}$               &7.63                     &2.79                    &10.1+11.5$i$             &20.5+11.5$i$      &1.2+0.673$i$       \\
$T^{(1)}_{K\Xi^*}$                    &3.81                     &1.39                    &-3.22+10.9$i$            &1.98+10.9$i$      &0.119+0.658$i$     \\
$T^{(0)}_{K\Xi^*}$                    &11.4                     &4.66                    &16.4+17.2$i$             &32.5+17.2$i$      &1.95+1.04$i$       \\
$T^{(1/2)}_{K\Omega}$                 &11.4                     &3.73                    &9.4+12.5$i$              &24.6+12.5$i$      &1.51+0.768$i$      \\
\hline
$T^{(2)}_{\bar K\Delta}$              &0                        &1.62                    &-2.85+8.34$i$            &-1.23+8.34$i$     &-0.0697+0.474$i$   \\
$T^{(1)}_{\bar K\Delta}$              &15.3                     &4.43                    &14.6+2.78$i$             &34.3+2.78$i$      &1.95+0.158$i$      \\
$T^{(3/2)}_{\bar K\Sigma^*}$          &-3.81                    &2.32                    &-4.01+11.3$i$            &-5.5+11.3$i$      &-0.323+0.662$i$    \\
$T^{(1/2)}_{\bar K\Sigma^*}$          &7.63                     &0.92                    &-0.775+2.42$i$           &7.77+2.42$i$      &0.456+0.142$i$     \\
$T^{(1)}_{\bar K\Xi^*}$               &-7.63                    &3.02                    &-5.33+8.51$i$            &-9.93+8.51$i$     &-0.598+0.512$i$    \\
$T^{(0)}_{\bar K\Xi^*}$               &0                        &-0.249                  &-9.72-0.533$i$           &-9.97-0.533$i$    &-0.6-0.0321$i$     \\
$T^{(1/2)}_{\bar K\Omega}$            &-11.4                    &3.73                    &-6.8                     &-14.5             &-0.892             \\
\hline
\end{tabular}
\end{table}

\begin{table}
\caption{Eta-light decuplet baryon threshold $T$-matrices order by order in unit of fm. }\label{etaT}
\begin{tabular}{cccccc}
\hline
                                      &$O(\epsilon^1)$          &$O(\epsilon^2)$         &$O(\epsilon^3)$          &Total             &Scattering Length  \\\hline
$T^{(3/2)}_{\eta\Delta}$              &0                        &-0.173                  &0.338+3.76$i$            &0.165+3.76$i$     &0.00909+0.207$i$   \\
$T^{(1)}_{\eta\Sigma^*}$              &0                        &3.58                    &1.79+9.96$i$             &5.37+9.96$i$      &0.306+0.568$i$     \\
$T^{(1/2)}_{\eta\Xi^*}$               &0                        &4.34                    &1.23+11.2$i$             &5.57+11.2$i$      &0.326+0.655$i$     \\
$T^{(0)}_{\eta\Omega}$                &0                        &2.1                     &-1.35+7.44$i$            &0.756+7.44$i$     &0.0453+0.446$i$    \\
\hline
\end{tabular}
\end{table}

In Ref. \cite{Bicudo2004}, the authors used the extended NJL model
and got the following $\pi\Delta$ scattering lengths:
 \begin{equation}
   a_{\pi \Delta}^{5/2}=-0.258 m_\pi^{-1}=-0.364~{\rm{fm}} , \quad
   a_{\pi \Delta}^{3/2}=0.172  m_\pi^{-1}= 0.242~{\rm{fm}} , \quad
   a_{\pi \Delta}^{1/2}=0.429  m_\pi^{-1}= 0.604~{\rm{fm}}.
 \end{equation}
Somehow their results are consistent with ours.

From the data, one easily sees that the third order contributions
are large and the $T$-matrices do not converge. The bad convergence
may be mainly due to the large mass of kaon and eta
\cite{Lee1994,Kaiser2001,Zhu2001b,Zhu2002}. Inaccurate
determination of LECs might worsen the convergence. Another reason
probably comes from the heavy hadron framework itself. In the
meson-baryon case \cite{Mai2009} and the meson-heavy meson case
\cite{Liu2009,Geng2010}, it was found that the recoil corrections
are sizable. The future investigation of the higher order recoil
corrections and LECs etc may answer whether similar features occur
in the meson-decuplet cases.

We also show different contributions of the $T$-matrices at
$O(\epsilon^3)$ in Table \ref{piT3}, \ref{KT3}, and \ref{etaT3}.
From the values one observes that the loop corrections from the
intermediate decuplet states dominate. In the pion scattering
channels, only loop diagrams containing the octet baryons could
generate the imaginary part. The contributions from the
intermediate octet and decuplet baryons have opposite signs in
many channels, which is beneficial for the convergence. The
consideration of the recoil corrections increases the convergence
in most channels.

We compare $T^{(2,1,0)}_{\pi\Sigma^*}$,
$T^{(3/2,1/2)}_{\pi\Xi^*}$, $T^{(3/2,1/2)}_{K\Sigma^*}$,
$T^{(1,0)}_{K\Xi^*}$, $T^{(3/2,1/2)}_{\bar K\Sigma^*}$ and
$T^{(1,0)}_{\bar K\Xi^*}$ with the corresponding
$T^{(2,1,0)}_{\pi\Sigma}$, $T^{(3/2,1/2)}_{\pi\Xi}$,
$T^{(3/2,1/2)}_{K\Sigma}$, $T^{(1,0)}_{K\Xi}$,
$T^{(3/2,1/2)}_{\bar K\Sigma}$ and $T^{(1,0)}_{\bar K\Xi}$
obtained in Refs. \cite{Liu2007,Liu2007a}, and find that the
results of each pair are equal at the leading order, but different
at higher order. We also check the following quantities at the
leading order,
\begin{equation}
  T_{\pi^+\Delta^+}=T_{\pi^+p},\quad  T_{\pi^+\Delta^0}=T_{\pi^+n}, \quad
  T_{K^+\Delta^+}= T_{K^+p},\quad  T_{K^+\Delta^0}=  T_{K^+n},
  \quad T_{\bar K^0 \Delta^+}=T_{\bar K^0 p},\quad T_{\bar K^0 \Delta^0}= T_{\bar K^0 n}.
\end{equation}
These relations are easy to understand with the Weinberg-Tomazawa
formula: $T\sim I_{tot}(I_{tot}+1)-I_B(I_B+1)-I_M(I_M+1)$ where
$I_{tot}$, $I_B$, and $I_M$ are the total, baryon, and meson
isospin, respectively.

In summary, we have calculated the S-wave scattering lengths of
the pseudoscalar mesons and light decuplet baryons to the third
order in the framework of heavy baryon chiral perturbation theory.
We estimate LECs with the resonance saturation method. Our results
may be helpful to the model construction of the meson-decuplet
baryon interactions. We hope the future lattice simulation or
experimental measurements may test the predictions.

\begin{table}
\caption{Pion-light decuplet baryon threshold $T$-matrices at $O(\epsilon^3)$ in unit of fm. }\label{piT3}
\begin{tabular}{cccccccccc}
\hline
                                                    &Loop i:                                       &Loop ii:                                            &Loop:                                         &Tree                                        &Total                 \\
                                                    &Octet Contribution                            &Decuplet  Contribution                              &Total                                         &Only Recoil Correction                      &                      \\\hline
$T^{(5/2)}_{\pi\Delta}$                             &0                                             &-1.41                                               &-1.41                                         &0.0345                                      &-1.37                 \\
$T^{(3/2)}_{\pi\Delta}$                             &-0.518-1.02$i$                                &-0.586                                              &-1.1-1.02$i$                                  &0.203                                       &-0.901-1.02$i$        \\
$T^{(1/2)}_{\pi\Delta}$                             &0.414+0.814$i$                                &1.08                                                &1.49+0.814$i$                                 &-0.0575                                     &1.44+0.814$i$         \\
$T^{(2)}_{\pi\Sigma^*}$                             &-0.129-0.102$i$                               &-2.41                                               &-2.54-0.102$i$                                &0.023                                       &-2.52-0.102$i$        \\
$T^{(1)}_{\pi\Sigma^*}$                             &-0.337-0.509$i$                               &-1.56                                               &-1.9-0.509$i$                                 &0.0977                                      &-1.8-0.509$i$         \\
$T^{(0)}_{\pi\Sigma^*}$                             &0.181+0.509$i$                                &-0.548                                              &-0.366+0.509$i$                               &-0.046                                      &-0.412+0.509$i$       \\
$T^{(3/2)}_{\pi\Xi^*}$                              &-0.129-0.102$i$                               &-2.57                                               &-2.7-0.102$i$                                 &0.0115                                      &-2.69-0.102$i$        \\
$T^{(1/2)}_{\pi\Xi^*}$                              &-0.129-0.102$i$                               &-1.78                                               &-1.91-0.102$i$                                &0.0222                                      &-1.89-0.102$i$        \\
$T^{(1)}_{\pi\Omega}$                               &0                                             &-1.87                                               &-1.87                                         &0                                           &-1.87                 \\
\hline
\end{tabular}
\end{table}

\begin{table}
\caption{Kaon-light decuplet baryon threshold $T$-matrices at $O(\epsilon^3)$ in unit of fm. }\label{KT3}
\begin{tabular}{cccccccccc}
\hline
                                                    &Loop i:                                       &Loop ii:                                            &Loop:                                         &Tree                                        &Total                 \\
                                                    &Octet Contribution                            &Decuplet  Contribution                              &Total                                         &Only Recoil Correction                      &                      \\\hline
$T^{(2)}_{K\Delta}$                                 &0                                             &-9.33                                               &-9.33                                         &0.0607                                      &-9.27                 \\
$T^{(1)}_{K\Delta}$                                 &0                                             &3.69                                                &3.69                                          &-0.0202                                     &3.67                  \\
$T^{(3/2)}_{K\Sigma^*}$                             &0.778-0.178$i$                                &-11.+6.95$i$                                        &-10.2+6.77$i$                                 &0.777                                       &-9.44+6.77$i$         \\
$T^{(1/2)}_{K\Sigma^*}$                             &0.666+0.355$i$                                &9.47+11.1$i$                                        &10.1+11.5$i$                                  &-0.0405                                     &10.1+11.5$i$          \\
$T^{(1)}_{K\Xi^*}$                                  &0.778-0.178$i$                                &-4.99+11.1$i$                                       &-4.21+10.9$i$                                 &0.989                                       &-3.22+10.9$i$         \\
$T^{(0)}_{K\Xi^*}$                                  &0.628+0.533$i$                                &15.8+16.7$i$                                        &16.4+17.2$i$                                  &-0.0607                                     &16.4+17.2$i$          \\
$T^{(1/2)}_{K\Omega}$                               &0                                             &8.7+12.5$i$                                         &8.7+12.5$i$                                   &0.696                                       &9.4+12.5$i$           \\
\hline
$T^{(2)}_{\bar K\Delta}$                            &0                                             &-2.85+8.34$i$                                       &-2.85+8.34$i$                                 &0                                           &-2.85+8.34$i$         \\
$T^{(1)}_{\bar K\Delta}$                            &0                                             &13.6+2.78$i$                                        &13.6+2.78$i$                                  &0.929                                       &14.6+2.78$i$          \\
$T^{(3/2)}_{\bar K\Sigma^*}$                        &0.703+0.178$i$                                &-4.73+11.1$i$                                       &-4.03+11.3$i$                                 &0.0202                                      &-4.01+11.3$i$         \\
$T^{(1/2)}_{\bar K\Sigma^*}$                        &0.816-0.355$i$                                &-3.06+2.78$i$                                       &-2.25+2.42$i$                                 &1.47                                        &-0.775+2.42$i$        \\
$T^{(1)}_{\bar K\Xi^*}$                             &0.703+0.178$i$                                &-6.07+8.34$i$                                       &-5.37+8.51$i$                                 &0.0405                                      &-5.33+8.51$i$         \\
$T^{(0)}_{\bar K\Xi^*}$                             &0.854-0.533$i$                                &-12.1                                               &-11.2-0.533$i$                                &1.51                                        &-9.72-0.533$i$        \\
$T^{(1/2)}_{\bar K\Omega}$                          &0                                             &-6.86                                               &-6.86                                         &0.0607                                      &-6.8                  \\
\hline
\end{tabular}
\end{table}

\begin{table}
\caption{Eta-light decuplet baryon threshold $T$-matrices at $O(\epsilon^3)$ in unit of fm. }\label{etaT3}
\begin{tabular}{cccccccccc}
\hline
                                                    &Loop i:                                       &Loop ii:                                            &Loop:                                         &Tree                                        &Total                 \\
                                                    &Octet Contribution                            &Decuplet  Contribution                              &Total                                         &Only Recoil Correction                      &                      \\\hline
$T^{(3/2)}_{\eta\Delta}$                            &0.493+0.0439$i$                               &-0.479+3.72$i$                                      &0.0145+3.76$i$                                &0.324                                       &0.338+3.76$i$         \\
$T^{(1)}_{\eta\Sigma^*}$                            &-0.548+0.0366$i$                              &2.33+9.92$i$                                        &1.79+9.96$i$                                  &0                                           &1.79+9.96$i$          \\
$T^{(1/2)}_{\eta\Xi^*}$                             &-0.403+0.0219$i$                              &1.3+11.2$i$                                         &0.902+11.2$i$                                 &0.324                                       &1.23+11.2$i$          \\
$T^{(0)}_{\eta\Omega}$                              &0.93                                          &-3.57+7.44$i$                                       &-2.64+7.44$i$                                 &1.29                                        &-1.35+7.44$i$         \\
\hline
\end{tabular}
\end{table}

\section*{Acknowledgments}

This project is supported by the National Natural Science
Foundation of China under Grants No. 10625521, No. 10721063, No. 10805048 and
the Ministry of Science and Technology of China (2009CB825200). YRL was supported by
the Japan Society for the Promotion of Science KAKENHI (Grant No. 21.09027).

\appendix

\section{Rough estimation of the $O(\epsilon^3)$ counter terms}\label{append}

At this order, the excited baryons contribute to LECs through both
tree diagrams and loop diagrams. Here we assume the former
contribution is much smaller than the latter one. This assumption
for the estimation is similar to the one used in studying the
meson-baryon scattering lengths in Refs. \cite{Kaiser2001,Liu2007}
and the above numerical calculation. For a very rough estimate at
$O(\epsilon^3)$, we only consider the octet resonance containing
$N(1440)$, denoted by $R$.

The diagrams one needs to calculate are similar to those in the last column of Fig. \ref{LoopDiag}.
A new Lagrangian should be introduced,
\begin{equation}
  {\mathcal L}_{\phi TR}=\Tr\left( \bar R (iv\cdot \partial -\delta_R) R+\bar R [v\cdot \Gamma, R]\right)
                      +G_{N(1440)} (\bar T^\mu u_\mu R +\bar R u_\mu T^\mu)+...
\end{equation}
where $\delta_R\sim500$ MeV is the mass difference between
$N(1440)$ and the nucleon. In our calculation, we need $\tilde
\delta_R=208$ MeV, which is the mass difference between $N(1440)$
and $\Delta$.

The $T$-matrices can be obtained by replacing $J$ and $W(m)$ in Eqs.
(\ref{J}) and (\ref{W}) with
\begin{equation}
  J_{\text{new}}=J+J_R, \qquad W_{\text{new}}(m)=W(m)+W_R(m),
\end{equation}
where
\begin{eqnarray}
  J_R&=&
  \frac{G_{N(1440)}^2 m_K^2}{12 \pi ^2 f^4 \left(m_\eta^2-m_\pi^2\right)}
     \left[\left( \tilde \delta_R ^3-\frac32 m_\eta^2 \tilde \delta_R \right) \log \frac{m_\eta}{\lambda}
     + (\frac32 m_\pi^2 \tilde \delta_R- \tilde \delta_R^3)\log\frac{m_\pi}{\lambda}\right.\nonumber\\
  &&\left.\qquad
     +( \tilde \delta_R ^2 -m_\pi^2)^{3/2} \log \frac{\sqrt{\tilde \delta_R ^2-m_\pi^2}-\tilde \delta_R }{m_\pi }
     -\left(m_\eta^2-\tilde \delta_R ^2\right)^{3/2} \cos ^{-1}\left(-\frac{\tilde \delta_R }{m_\eta}\right)\right.
        \nonumber\\
  &&\left. \qquad
     +(m_\eta^2 - m_\pi^2) \tilde \delta_R    \frac{}{}
     \right], \\
  W_R(m)&=&  \frac{G_{N(1440)}^2}{16 \pi ^2 f^4}
  \begin{cases}
     -2 \sqrt{m^2-\tilde \delta_R ^2} \cos ^{-1}\left(-\frac{\tilde \delta_R }{m}\right)
     -2 \tilde \delta_R  \log\frac{m}{\lambda }+\tilde \delta_R                 & m>|\tilde \delta_R|  \\
     -2\tilde \delta_R\log\frac{m}{\lambda }
     +2\sqrt{\tilde \delta_R ^2-m^2}\log\frac{\sqrt{\tilde \delta_R ^2-m^2}-\tilde \delta_R }{m }
     +\tilde \delta_R                                                          & m\leq |\tilde \delta_R|
  \end{cases}    .
\end{eqnarray}

We show the numerical results at order $O(\epsilon^3)$ in Tables \ref{piTN}, \ref{KTN} and \ref{etaTN}. The
roughly estimated LECs are $h_1^r=-1.7$ GeV$^{-2}$, $h_2^r=-4.4$ GeV$^{-2}$, $h_3^r=3.3$ GeV$^{-2}$,
$h_4^r=-0.8$ GeV$^{-2}$, and $h_5^r=0.0$ at around the scale $4\pi f_\pi$. From the tables, we see that the
contribution due to the intermediate $N(1440)$ octet increases the convergence of the chiral expansion in most
channels slightly.

\begin{table}
\caption{Pion-light decuplet baryon threshold $T$-matrices considering $N(1440)$ octet baryon contribution
                in unit of fm. }\label{piTN}
\begin{tabular}{cccccc}
\hline
                                 &$O(\epsilon^1)$   &$O(\epsilon^2)$  &$O(\epsilon^3)$ i:                  &$O(\epsilon^3)$ ii:                &$O(\epsilon^3)$:   \\
                                 &                  &                 &\qquad\qquad Old~\qquad\qquad\qquad & $N(1440)$ Contribution            &~~~~Total, New~~~~ \\\hline
$T^{(5/2)}_{\pi\Delta}$          &-4.84             &0.445            &-1.37                               &0                                  &-1.37              \\
$T^{(3/2)}_{\pi\Delta}$          &3.23              &-0.253           &-0.901-1.02$i$                      &0.374                              &-0.527-1.02$i$     \\
$T^{(1/2)}_{\pi\Delta}$          &8.06              &1.               &1.44+0.814$i$                       &-0.299                             &1.14+0.814$i$      \\
$T^{(2)}_{\pi\Sigma^*}$          &-3.23             &0.361            &-2.52-0.102$i$                      &-0.0434                            &-2.56-0.102$i$     \\
$T^{(1)}_{\pi\Sigma^*}$          &3.23              &0.0819           &-1.8-0.509$i$                       &0.106                              &-1.69-0.509$i$     \\
$T^{(0)}_{\pi\Sigma^*}$          &6.45              &0.781            &-0.412+0.509$i$                     &-0.268                             &-0.68+0.509$i$     \\
$T^{(3/2)}_{\pi\Xi^*}$           &-1.61             &0.277            &-2.69-0.102$i$                      &-0.0434                            &-2.73-0.102$i$     \\
$T^{(1/2)}_{\pi\Xi^*}$           &3.23              &0.277            &-1.89-0.102$i$                      &-0.0434                            &-1.93-0.102$i$     \\
$T^{(1)}_{\pi\Omega}$            &0                 &0.194            &-1.87                               &0                                  &-1.87              \\
\hline
\end{tabular}
\end{table}

\begin{table}
\caption{Kaon-light decuplet baryon threshold $T$-matrices considering $N(1440)$ octet baryon contribution
                in unit of fm. }\label{KTN}
\begin{tabular}{cccccc}
\hline
                                 &$O(\epsilon^1)$   &$O(\epsilon^2)$  &$O(\epsilon^3)$ i:                  &$O(\epsilon^3)$ ii:                &$O(\epsilon^3)$:   \\
                                 &                  &                 &\qquad\qquad Old~\qquad\qquad\qquad & $N(1440)$ Contribution            &~~~~Total, New~~~~ \\\hline
$T^{(2)}_{K\Delta}$              &-11.4             &3.73             &-9.27                               &0                                  &-9.27              \\
$T^{(1)}_{K\Delta}$              &3.81              &0.917            &3.67                                &0                                  &3.67               \\
$T^{(3/2)}_{K\Sigma^*}$          &-3.81             &1.39             &-9.44+6.77$i$                       &1.62                               &-7.82+6.77$i$      \\
$T^{(1/2)}_{K\Sigma^*}$          &7.63              &2.79             &10.1+11.5$i$                        &-0.558                             &9.53+11.5$i$       \\
$T^{(1)}_{K\Xi^*}$               &3.81              &1.39             &-3.22+10.9$i$                       &1.62                               &-1.6+10.9$i$       \\
$T^{(0)}_{K\Xi^*}$               &11.4              &4.66             &16.4+17.2$i$                        &-1.28                              &15.1+17.2$i$       \\
$T^{(1/2)}_{K\Omega}$            &11.4              &3.73             &9.4+12.5$i$                         &0                                  &9.4+12.5$i$        \\
\hline
$T^{(2)}_{\bar K\Delta}$         &0                 &1.62             &-2.85+8.34$i$                       &0                                  &-2.85+8.34$i$      \\
$T^{(1)}_{\bar K\Delta}$         &15.3              &4.43             &14.6+2.78$i$                        &0                                  &14.6+2.78$i$       \\
$T^{(3/2)}_{\bar K\Sigma^*}$     &-3.81             &2.32             &-4.01+11.3$i$                       &0.168                              &-3.84+11.3$i$      \\
$T^{(1/2)}_{\bar K\Sigma^*}$     &7.63              &0.92             &-0.775+2.42$i$                      &2.35                               &1.57+2.42$i$       \\
$T^{(1)}_{\bar K\Xi^*}$          &-7.63             &3.02             &-5.33+8.51$i$                       &0.168                              &-5.16+8.51$i$      \\
$T^{(0)}_{\bar K\Xi^*}$          &0                 &-0.249           &-9.72-0.533$i$                      &3.07                               &-6.65-0.533$i$     \\
$T^{(1/2)}_{\bar K\Omega}$       &-11.4             &3.73             &-6.8                                &0                                  &-6.8               \\
\hline
\end{tabular}
\end{table}

\begin{table}
\caption{Eta-light decuplet baryon threshold $T$-matrices considering $N(1440)$ octet baryon contribution
                in unit of fm. }\label{etaTN}
\begin{tabular}{cccccc}
\hline
                                 &$O(\epsilon^1)$   &$O(\epsilon^2)$  &$O(\epsilon^3)$ i:                  &$O(\epsilon^3)$ ii:                &$O(\epsilon^3)$:   \\
                                 &                  &                 &\qquad\qquad Old~\qquad\qquad\qquad & $N(1440)$ Contribution            &~~~~Total, New~~~~ \\\hline
$T^{(3/2)}_{\eta\Delta}$         &0                 &-0.173           &0.338+3.76$i$                       &0.814                              &1.15+3.76$i$       \\
$T^{(1)}_{\eta\Sigma^*}$         &0                 &3.58             &1.79+9.96$i$                        &-0.686                             &1.1+9.96$i$        \\
$T^{(1/2)}_{\eta\Xi^*}$          &0                 &4.34             &1.23+11.2$i$                        &-0.402                             &0.823+11.2$i$      \\
$T^{(0)}_{\eta\Omega}$           &0                 &2.1              &-1.35+7.44$i$                       &1.67                               &0.319+7.44$i$      \\
\hline
\end{tabular}
\end{table}

\end{document}